\documentclass[
               biblatex,     
               ]{jacow}

\usepackage[english]{babel}

\begin{document}

\title{HIGH-POWER TEST AND SYSTEM INTEGRATION OF DIRECT RF SAMPLING BASED LLRF CONTROL AND MONITORING SYSTEM FOR S-BAND ACCELERATING STRUCTURES\thanks{This work was supported by the U.S. DOE, Office of Science contract DE-AC02-76SF00515.}}

\author{C.~Liu\thanks{chaoliu@slac.stanford.edu}, A.~Dhar, M.~Hoganson, J.~Olszewski, E.~Snively, W.~H.~Tan, S.~Morton, T.~Le, E.~Nanni\\SLAC National Accelerator Laboratory, Menlo Park, United States}

\maketitle


\begin{abstract}
High precision Low-level RF (LLRF) control and monitoring systems for future particle accelerators will be a significant technical challenge as the requirements in performance, flexibility and affordability become increasingly stringent. We have developed an RF system-on-chip (RFSoC) based next-generation LLRF (NG-LLRF) for S-band accelerating structures, which samples and synthesizes the RF pulses directly without the analog mixers used for traditional LLRF systems. The platform delivered considerably better performance than the requirements of the targeted applications, such as the upgrades for Next Linear Collider Test Accelerator (NLCTA) and test facilities at SLAC. As part of the upgrade program, we also developed a custom solid-state amplifier (SSA) to deliver RF pulses at the desired power level of the klystron. Integration of the LLRF with the SSA and the high-power test facility could be challenging. The power levels and RF pulse stability at each stage of the high-power RF drive system must be optimized to deliver the desired RF performance. In this paper, the integration procedure and the test and characterization results at each stage of integration will be summarized, analyzed and discussed. This integration is an essential step for the full deployment of the NG-LLRF system to test facilities and accelerators in different frequency bands.
\end{abstract}

\section{Introduction}

Direct RF sampling technology has been applied to a range of astrophysics \cite{liu2021characterizing, liu2022development, henderson2022advanced, kurtz2022born, liu2023evaluating,liu2024development}and high energy physics experiments and projects \cite{einstein2023llrf,liu:ipac2024-mocn2, liu2025compact, liu2025higha,liu:ipac2025-thps140,liu2024next} at SLAC and the wider community. In \cite{liu2025high}, the results of a high-power test of NG-LLRF with a prototype C-band accelerator were summarized, remarking the successful integration of an NG-LLRF and a high-power test stand. Based on research and development for C-band accelerators, we adapt the operating frequency of the NG-LLRF system to S-band and the RF stability demonstrated with a loop back test setup is considerably better than the requirements of LCLS \cite{liu:ipac2025-thps141}. We began to integrate S-band NG-LLRF and a new SSA with test facilities at SLAC, starting from NLCTA S-band test station. with synergy of the research in Artificial Intelligence (AI) and Machine Learning (ML) based autonomous control for particle accelerator, we design and implement the S-band NG-LLRF as a base platform for AI/ML ready dataset accumulation and edge deployment of real-time control algorithms. The NG-LLRF has distinctive advantages compared with conventional heterodyne based LLRF that are essential for AI/ML based control, including wider RF bandwidth coverage, higher flexibility, higher data streaming throughput, and larger integrated computation resources. In this paper, the initial integration stages between NG-LLRF and NLCTA are described, summarized, and discussed.      

\section{System Integration}

Integration of the NG-LLRF and the custom SSA with the S-band station has been performed  stage by stage. In this section, the key components will be described and  the characterization results will be summarized.

\subsection{S-band NG-LLRF}

\begin{figure}[!htb]
   \centering
   \includegraphics*[width=\columnwidth]{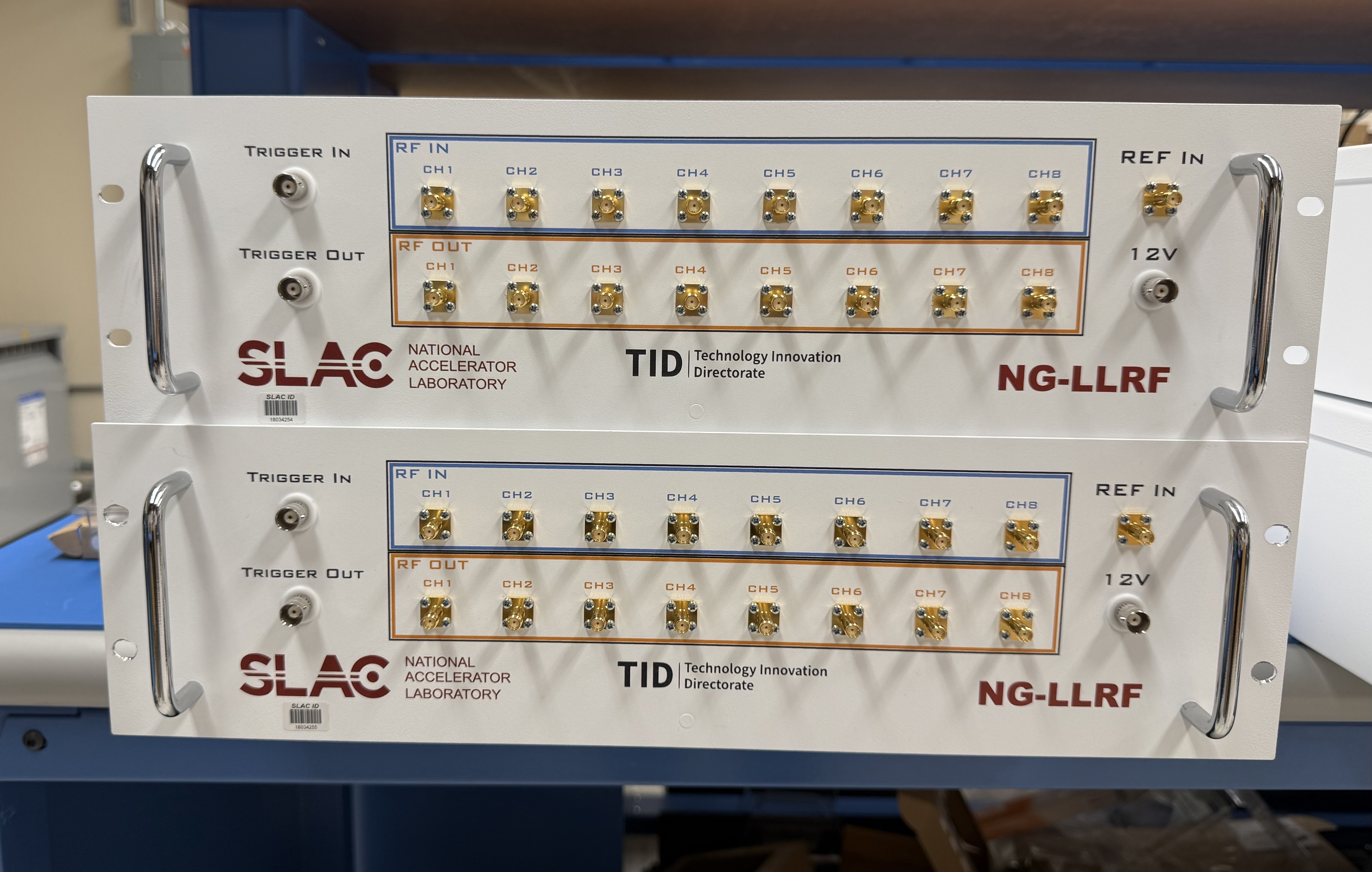}
   \caption{The front panel of the NG-LLRF chassis with 8 RF inputs and 8 RF outputs.}
   \label{fig:ngllrf}
\end{figure}

The NG-LLRF chassis for S-band has been built based on the ZCU208 evaluation board \cite{zcu208}. Figure \ref{fig:ngllrf} shows the front panel of the chassis. The chassis has 8 RF input channels with sampling rate up to 5 giga samples per second (GSPS), which has a maximum RF input frequency of 6 GHz, and 8 RF output channels with sampling rate up to 7 GSPS with current system configuration. The NG-LLRF can be triggered by an external trigger signal or be used as a master trigger source for other components, such SSA, klystron and etc. The NG-LLRF can be clocked from an internal oscillator or locked to an external reference using the "REF IN" port on the front panel. The control and data streaming interfaces are on the back plane, including GbE for remote update of the firmware and software images, control, configuration and low speed data streaming, and two SFP connectors for timing and high-speed data streaming. 

\subsection{Custom Solid State Amplifier (SSA)}

\begin{figure}[!htb]
   \centering
   \includegraphics*[width=\columnwidth]{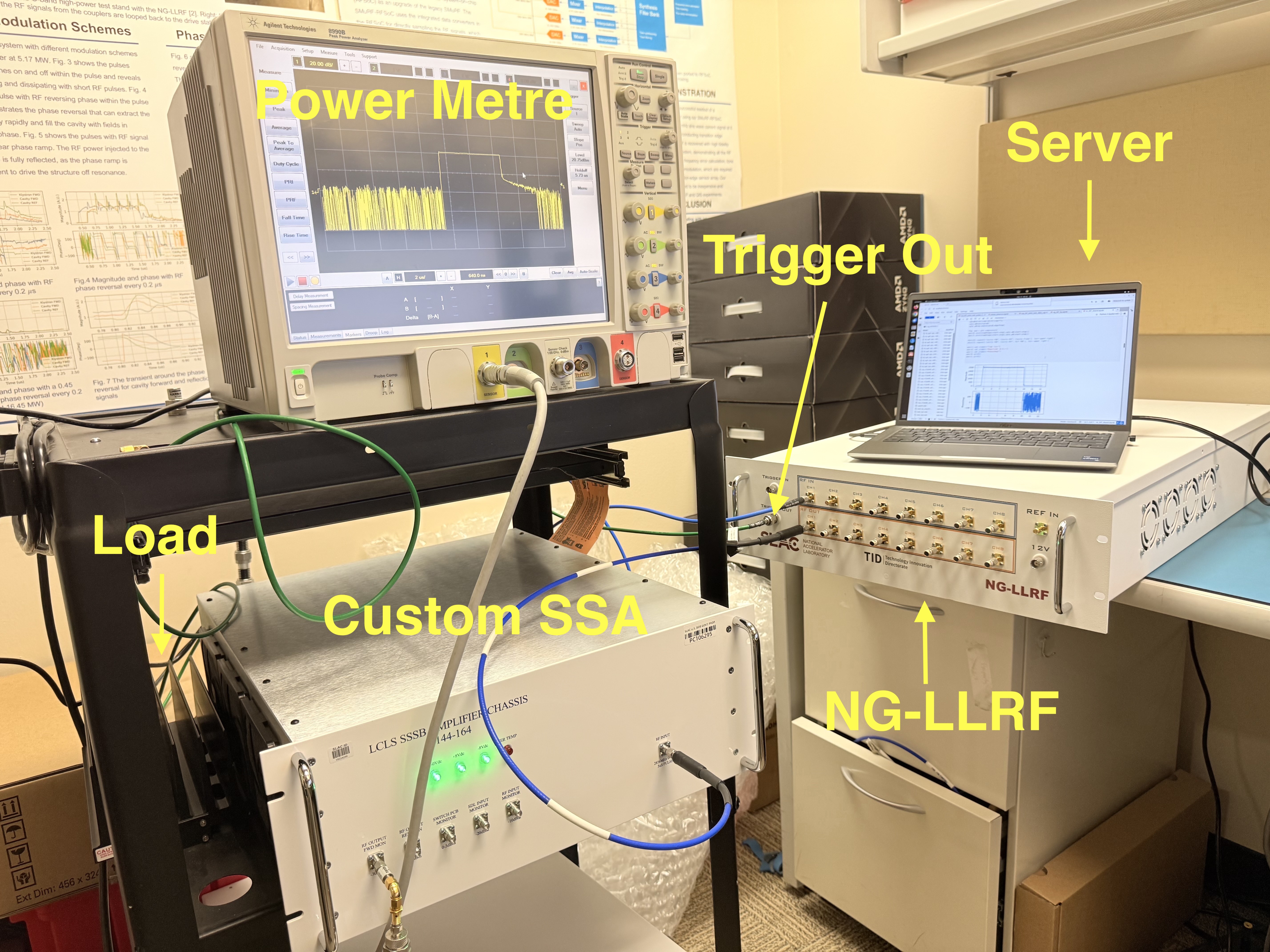}
   \caption{The setup for testing the peak RF power of SSA.}
   \label{fig:ssapwr}
\end{figure}

The custom SSA chassis has been built with a commercial SSA module and custom designed power supply boards. The SSA module has a rated peak RF power of 1 kW, maximum pulse width of 10 \(\mu\)s and maximum duty cycle 0.1\%. Figure \ref{fig:ssapwr} shows the setup to testing the peak SSA power using the NG-LLRF as RF drive. The NG-LLRF has been configured to generate RF pulses with pulse width up to 10 \(\mu\)s at 120 Hz. The power supply of the SSA is enabled by a trigger pulse generated by NG-LLRF, and the power of the SSA is enabled for more than 5 \(\mu\)s before the RF pulse arrives for the SSA to ramp up. The output of SSA has been connected to an RF load via an RF coupler, where the output power can be measured with appropriate attenuation. The RF output power can also be measured from an RF monitoring port on the SSA chassis, which is connected the RF coupler in the SSA module with a known attenuation. In this test setup, the monitoring output has been further attenuated and connected a peak power metre for RF power measurement.

\begin{figure}[!htb]
   \centering
   \includegraphics*[width=\columnwidth]{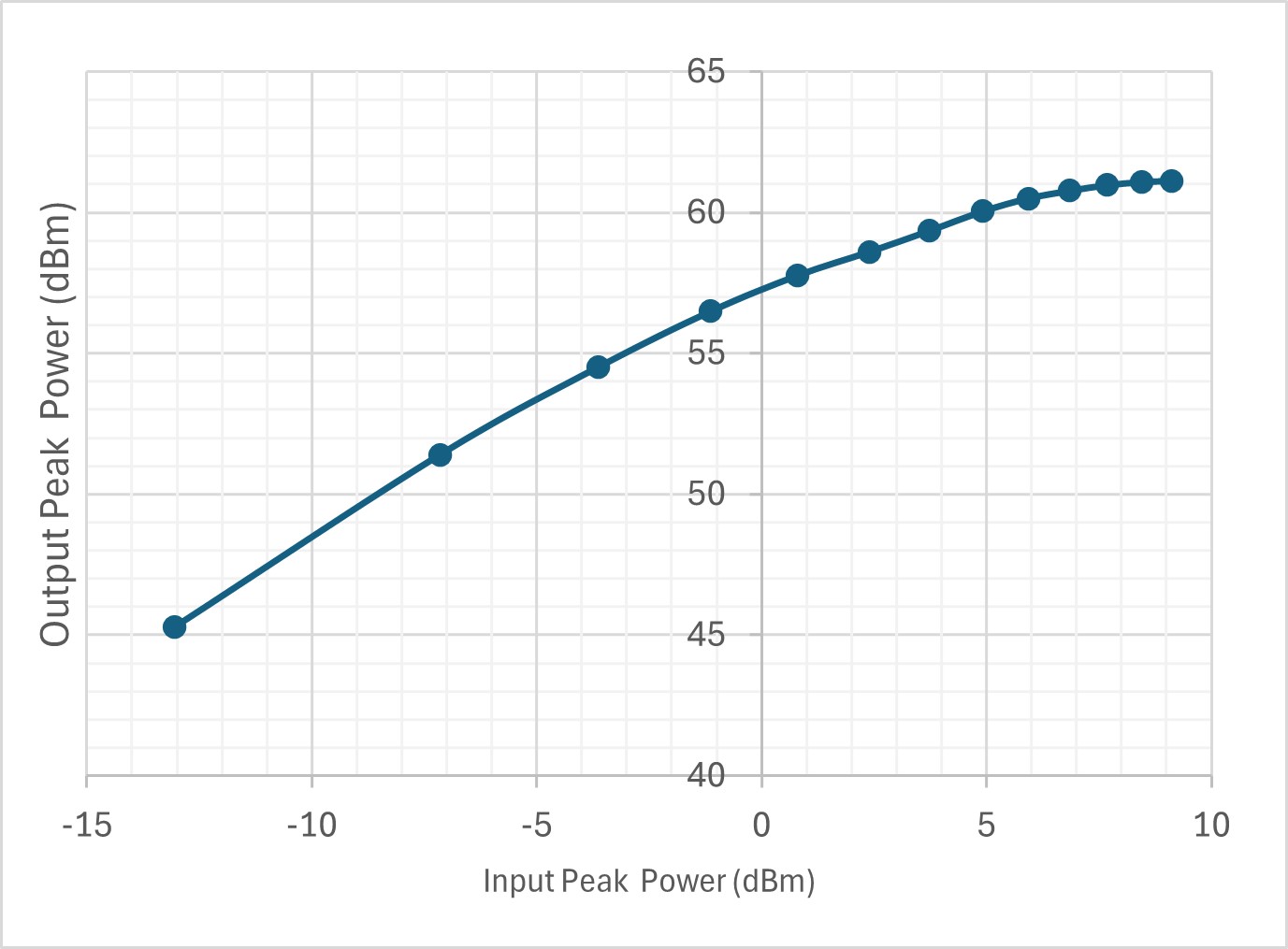}
   \caption{The input RF power versus output RF power of the SSA. }
   \label{fig:f3}
\end{figure}

Figure \ref{fig:f3} shows the input power versus output power measured in the input power range from -15.05 to 9.12 dBm. The output achieved 60 dBm, which is equivalent to approximately 1 kW and the requirement of NLCTA S-band klystron, at input level of 4.9 dBm. The highest power achieved is 61.12 dBm, which is equivalent to 1294 W. The SSA has around 58 dB gain in lower input levels and the gain reduced to 52 dB after saturation.  

\subsection{SSA and NG-LLRF}

The SSA and NG-LLRF have been fully integrated to perform a loop-back test. The NG-LLRF has been configured to generate RF pulses at 120 Hz with an RF frequency of 2856 MHz, which drives the SSA. The SSA output monitoring port, which as an integrated 30 dB attenuation to the SSA output, has been looped back to an RF input channel of NG-LLRF via another 10 dB attenuator. The RF input channel directly digitizes the RF signal and then down mixes the samples at 2856 MHz in the digital domain. The baseband pulses are streamed to the host server and stored for further analysis.

\begin{figure}[!htb]
   \centering
   \includegraphics*[width=\columnwidth]{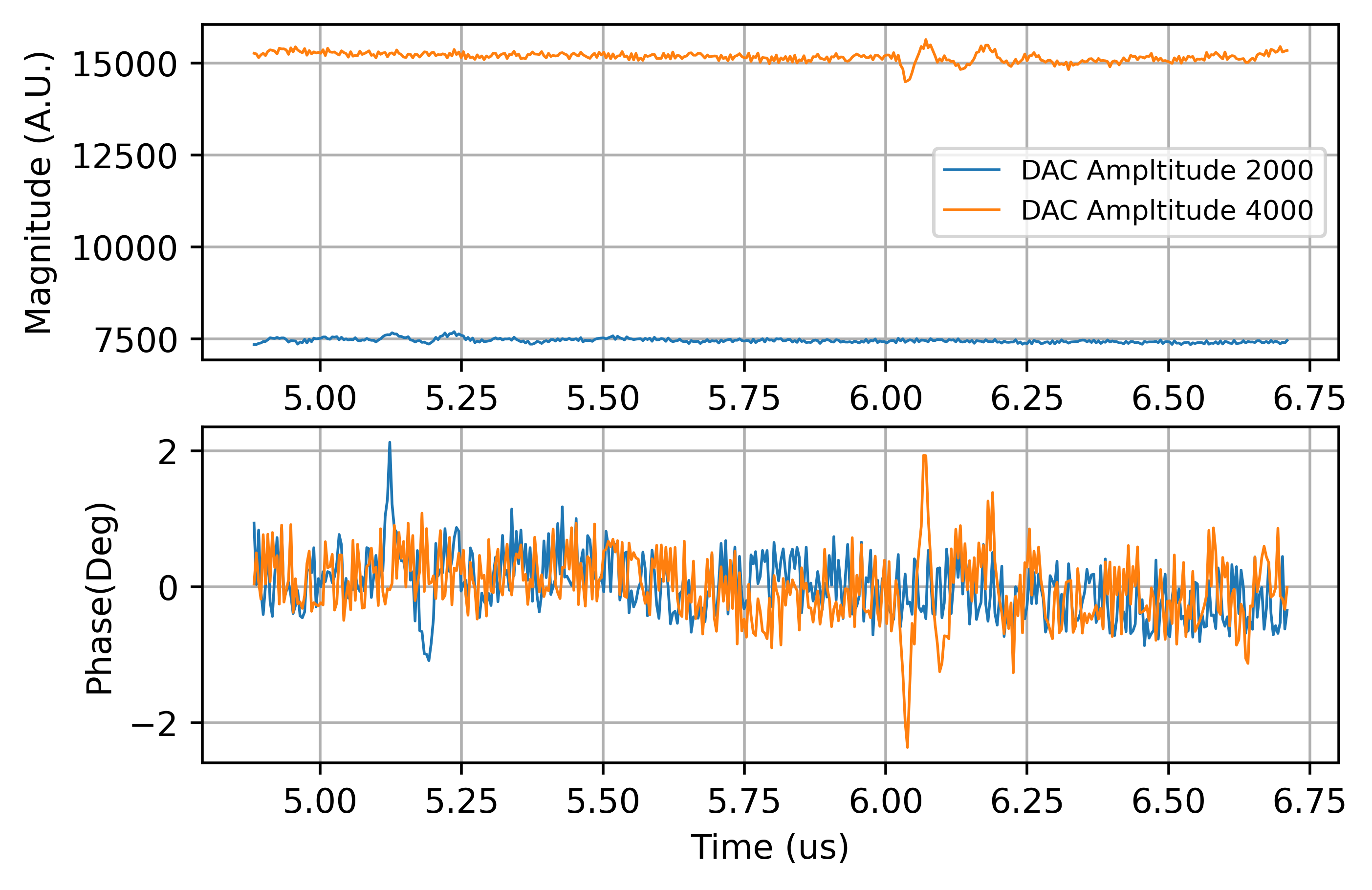}
   \caption{Pulse top fluctuation with 2 \(\mu\)s wide pulse at two different input power levels. }
   \label{fig:f4}
\end{figure}

\begin{figure}[!htb]
   \centering
   \includegraphics*[width=\columnwidth]{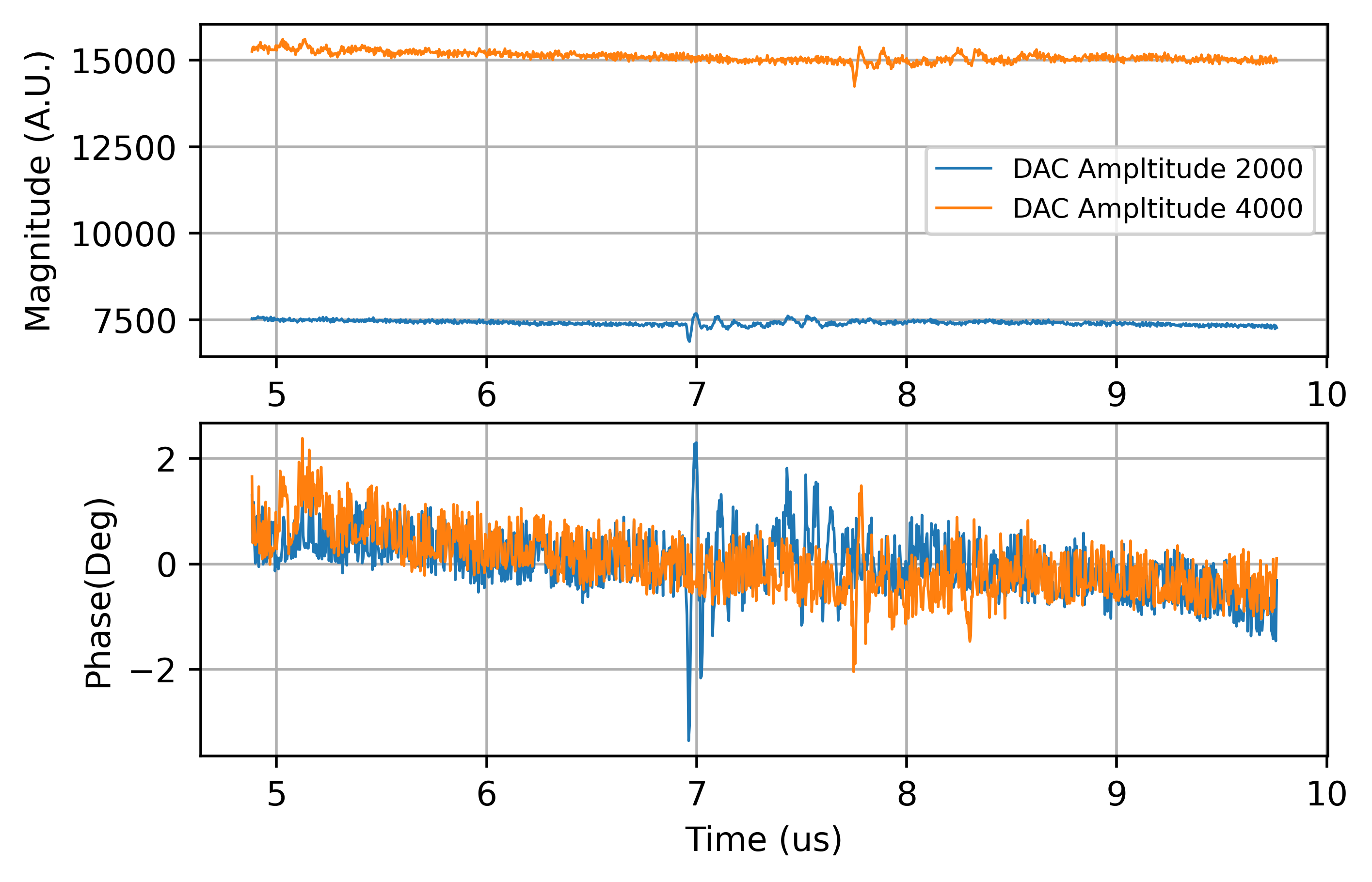}
   \caption{Pulse top fluctuation with 5 \(\mu\)s wide pulse at two different input power levels.}
   \label{fig:f5}
\end{figure}

NG-LLRF has been configured to generate RF pulses with durations of 2 \(\mu\)s and  5 \(\mu\)s at different power levels. Figure 4 shows the pulse top of the SSA output RF pulse captured with 2 \(\mu\)s. The traces are label with DAC amplitude levels, which sets the RF pulse amplitude in NG-LLRF software. In this case, DAC amplitude of 2000 generates RF pulse with peak power approximately -13.05 dBm, and DAC amplitude of 4000 generates peak power around -7.15 dBm. The amplitude of SSA output at DAC amplitude 4000 is almost exactly twice the amplitude of DAC amplitude 2000 shown in Figure 4, indicating that SSA operates with high linearity in this input power range. The SSA can be operated in higher DAC values with higher attenuation to utilize the full range of the DAC. There are oscillations on both amplitude and phase traces, which might be the result of mismatch between the load and the SSA causing power to reflect back and forth. Similar oscillations can be observed in multiple locations in the 5 \(\mu\)s pulse top plots shown in Figure \ref{fig:f5}. For applications that have stringent pulse top flatness requirements, the oscillations need to be further investigated and resolved. 

\begin{figure}[!htb]
   \centering
   \includegraphics*[width=\columnwidth]{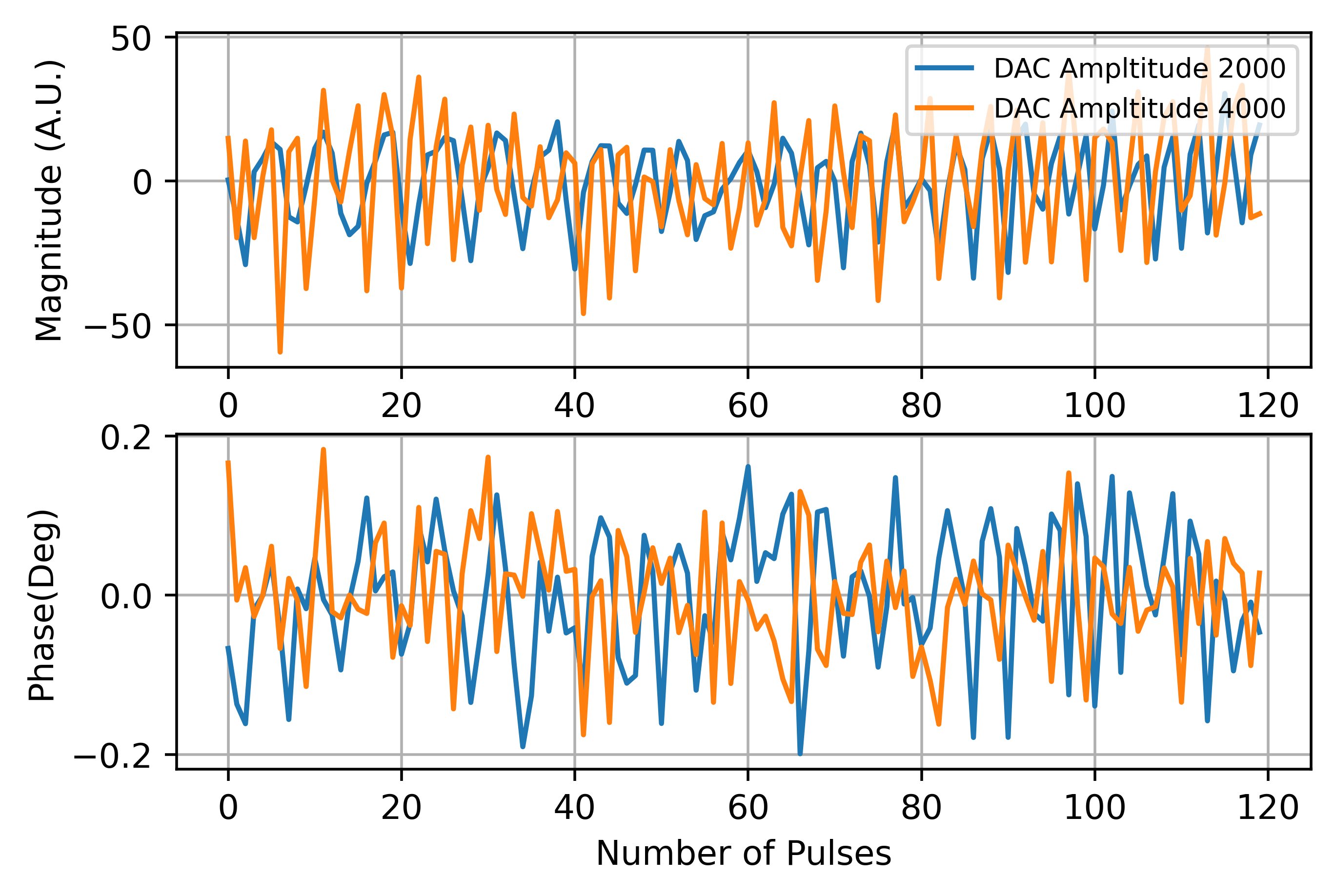}
   \caption{Pulse to pulse fluctuation of 120 consecutive 2 \(\mu\)s wide pulses at two different input power levels. }
   \label{fig:f6}
\end{figure}

\begin{figure}[!htb]
   \centering
   \includegraphics*[width=\columnwidth]{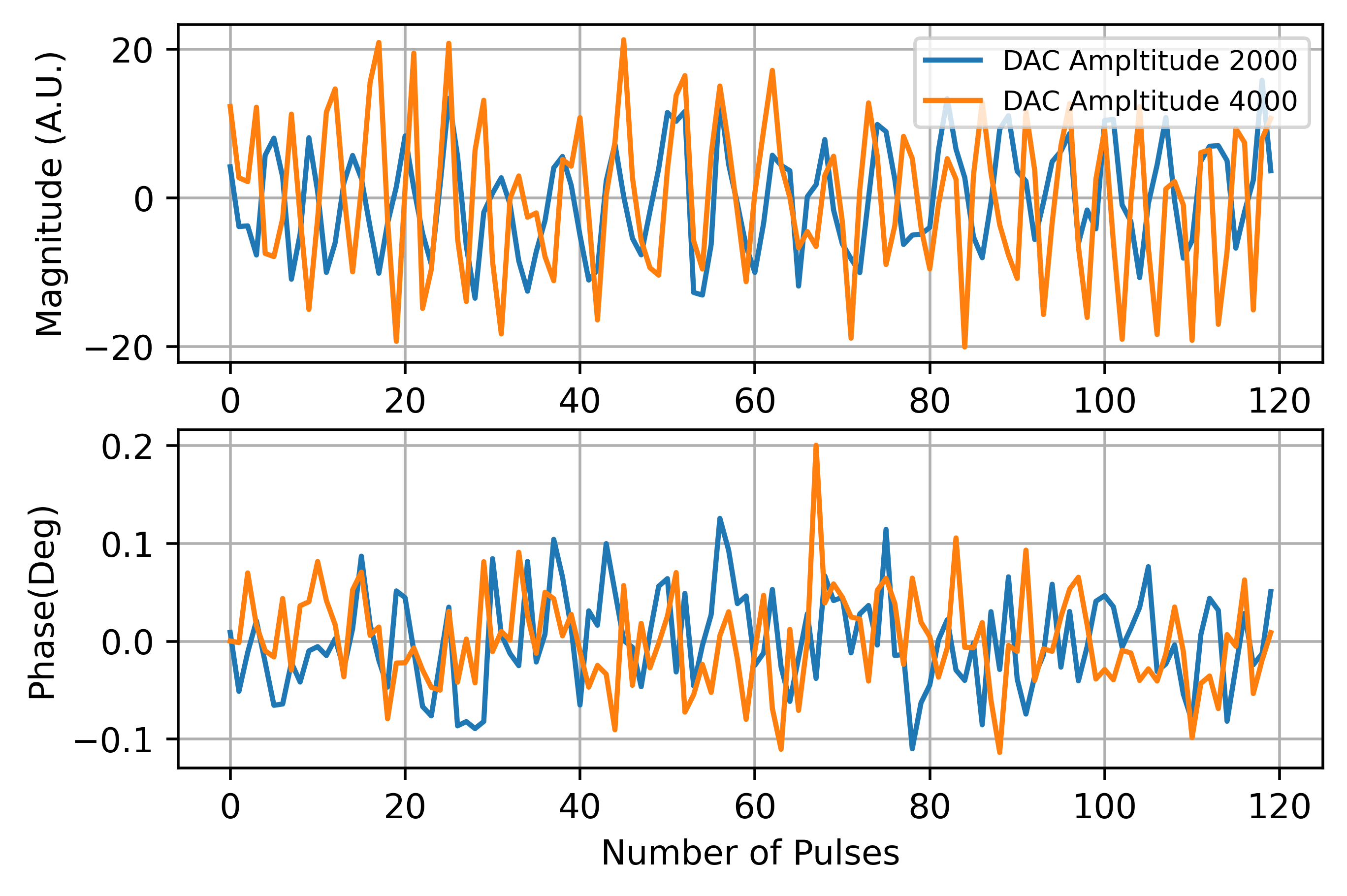}
   \caption{Pulse to pulse fluctuation of 120 consecutive 5 \(\mu\)s wide pulses at two different input power levels.}
   \label{fig:f7}
\end{figure}

Pulse to pulse fluctuation levels have been studied with 120 consecutive pulses captured in two different input power levels. Figure \ref{fig:f6} and \ref{fig:f7} show the average amplitude and phase of the pulses 2 and 5 \(\mu\)s, respectively. The mean values have been subtracted for easier visualization. The fluctuations in both amplitude and phase with 2 \(\mu\)s pulses have a higher frequency and higher absolute value than 5 \(\mu\)s pulses.

The pulse to pulse fluctuation levels have been further studied by calculation of the standard deviations. For amplitude, the ratio of the standard deviation of the average values of all 120 pulses and the mean of all the average amplitude values measured in percentage is used for evaluation. For phase, the standard deviation of all the average phase values of 120 pulses is directly used for evaluation. Table \ref{tab:t1} summarized the fluctuation levels with input power at -13.05 dBm. The amplitude fluctuation level with 5 \(\mu\)s pulses is about 50\% lower than 2 \(\mu\)s pulses, and the phase fluctuation is approximately 42\% lower. The improvement can mainly be attributed to the longer averaging time within each of the pulses. Table \ref{tab:t2} summarized the fluctuation levels with input power at -7.15 dBm. The amplitude fluctuation level with input power at -7.15 dBm is approximately 25\% lower than input power 6 dB lower. However, the phase fluctuation at higher power is not substantially lower, especially with 5 \(\mu\)s pulses. An amplitude fluctuation level lower than 0.1\% and phase fluctuation level lower than 0.05\textdegree are generally sufficient for most linear particle accelerators \cite{c3_2025esppu}. The tests are performed in open loop setup, and we are developing the closed loop system. 

\begin{table}[!hbt]
   \centering
   \caption{Amplitude and phase fluctuation levels with 2 \(\mu\)s and 5 \(\mu\)s wide pulses at 120 Hz in 1s with DAC amplitude at 2000 (equivalent to -13.05 dBm input power).}
   \begin{tabular}{lcc}
       \toprule
       \textbf{Width (\(\mu\)s)}  & \textbf{Amp (\%)}  & \textbf{Phase(\textdegree)} \\
       \midrule
           2       & 0.190           & 0.085        \\ 
           5       & 0.094           & 0.049         \\ 
       \bottomrule
   \end{tabular}
   \label{tab:t1}
\end{table}

\begin{table}[!hbt]
   \centering
   \caption{Amplitude and phase fluctuation levels with 2 \(\mu\)s and 5 \(\mu\)s wide pulses at 120 Hz in 1s with DAC amplitude at 4000 (equivalent to -7.15 dBm input power). }
   \begin{tabular}{lcc}
       \toprule
       \textbf{Width (\(\mu\)s)}  & \textbf{Amp (\%)}  & \textbf{Phase(\textdegree)} \\
       \midrule
           2       & 0.139           & 0.072        \\ 
           5       & 0.067           & 0.048         \\ 
       \bottomrule
   \end{tabular}
   \label{tab:t2}
\end{table}

\section{CONCLUSION}

This paper summarized the progress we have made to date in integrating the NG-LLRF and custom SSA to the S-band test station at NLCTA. We have used the NG-LLRF system to capture RF signals for high power tests with existing drive and it has demonstrated high precision RF measurement capabilities \cite{liu:icalepcs2025-modr005}. For the next step, we will integrate the triggering of NG-LLRF with the triggering system of NLCTA and eventually use the NG-LLRF and the custom SSA as the drive of klystron. 

The highly configurable digital RF front-end, high throughput data stream interfaces, and the massive real-time computation resources of NG-LLRF system make it an ideal platform for large scale AI/ML ready dataset accumulation and edge deployment experiments. The high resolution RF waveforms captured at different types of RF stations and beam instrumentation with consistent timestamp and format will be the foundation for training, verifying and optimizing large autonomous control algorithms. NLCTA with NG-LLRF can potentially be used as a testbed for future large scale AI/ML control model training and deployments. 

\printbibliography

\end{document}